\PassOptionsToPackage{table}{xcolor}
\documentclass[sigconf]{acmart} 
\usepackage{colortbl} % row colors
\usepackage[capitalise]{cleveref}
\usepackage[most]{tcolorbox} % for styled boxes

\tcbset{
  interviewquote/.style={
    enhanced,
    breakable,
    colback=white,        % light blue background
    colframe=white,      % frame color (light blue)
    coltitle=white,
    left=1em,
    right=1em,
    top=0.5em,
    bottom=0.5em,
    boxrule=0pt,           % no visible frame
    borderline west={2pt}{0pt}{purple!60}, % colored bar on the left
    sharp corners,
    fonttitle=\bfseries,
  }
}

\usepackage{framed}
\usepackage{changepage}

\newcommand{\interviewquote}[2]{
 \def\FrameCommand{%
    \hspace{0pt}%
    {\color{orange} \vrule width 2pt}% <-- Change color here.
    \colorbox{white}
  }%
  \MakeFramed{\advance\hsize-\width\FrameRestore}%
  \noindent% disable indenting first paragraph
  \begin{adjustwidth}{}{1pt}
  {\small``\textit{#1}'' - {#2}}\end{adjustwidth}\endMakeFramed%
}

\copyrightyear{2026}
\acmYear{2026}
\setcopyright{cc}
\setcctype{by-nc-nd}
\acmConference[ICSE-SEET '26]{2026 IEEE/ACM 48th International Conference on Software Engineering}{April 12--18, 2026}{Rio de Janeiro, Brazil}
\acmBooktitle{2026 IEEE/ACM 48th International Conference on Software Engineering (ICSE-SEET '26), April 12--18, 2026, Rio de Janeiro, Brazil}
\acmPrice{}
\acmDOI{10.1145/3786580.3786983}
\acmISBN{979-8-4007-2423-7/2026/04}

\begin{document}

%\title{The First Generation of AI-Assisted Programming Learners: Gendered Patterns in Critical Thinking and AI Ethics of German Secondary School Students}
\title[Learning to Program Alongside AI]{Learning to Program Alongside AI: Critical Thinking, AI Ethics, and Gendered Patterns of German Secondary School Students}

\author{Isabella Gra{\ss}l}
\email{isabella.grassl@tu-darmstadt.de}
\orcid{0000-0001-5522-7737}
\affiliation{%
  \institution{Technical University of Darmstadt}
  \city{Darmstadt}
  \country{Germany}}

\begin{abstract}
The first generation of students is learning to program alongside GenAI (\emph{Generative Artificial Intelligence}) tools, raising questions about how young learners critically engage with them and perceive ethical responsibilities. While prior research has focused on university students or developers, little is known about secondary school novices, who represent the next cohort of software engineers.
To address this gap, we conducted an exploratory study with 84 German secondary school students aged 16--19 attending software development workshops. We examined their critical thinking practices in AI-assisted programming, perceptions of AI ethics and responsibility, and gender-related differences in their views. 
Our results reveal an AI paradox: students demonstrate strong ethical reasoning and awareness about AI, yet many report integrating AI-generated code without a thorough understanding of it. The majority of our cohort attributed significant responsibility for AI practices to politics and corporations, potentially reflecting Germany's cultural context, with its strict regulations and data privacy discourse. Boys reported more frequent and experimental use of AI-assisted programming, whereas girls expressed greater scepticism and emphasised peer collaboration over GenAI assistance.  
Our findings highlight the importance of culturally responsive software engineering education that strengthens critical AI literacy in AI-assisted programming by linking ethics to concrete code artefacts and preparing young learners for this AI-driven software landscape.
\end{abstract}

\begin{CCSXML}
<ccs2012>
   <concept>
       <concept_id>10003456.10003457.10003527.10003531.10003533</concept_id>
       <concept_desc>Social and professional topics~Computer science education</concept_desc>
       <concept_significance>500</concept_significance>
       </concept>
   <concept>
       <concept_id>10003456.10010927</concept_id>
       <concept_desc>Social and professional topics~User characteristics</concept_desc>
       <concept_significance>500</concept_significance>
       </concept>
   <concept>
       <concept_id>10010405.10010489</concept_id>
       <concept_desc>Applied computing~Education</concept_desc>
       <concept_significance>500</concept_significance>
       </concept>
   <concept>
       <concept_id>10010147.10010178</concept_id>
       <concept_desc>Computing methodologies~Artificial intelligence</concept_desc>
       <concept_significance>500</concept_significance>
       </concept>
 </ccs2012>
\end{CCSXML}

\ccsdesc[500]{Social and professional topics~Computer science education}
\ccsdesc[500]{Social and professional topics~User characteristics~Gender}
\ccsdesc[500]{Applied computing~Education}
\ccsdesc[500]{Computing methodologies~Artificial intelligence}

\keywords{Programming, GenAI, AI Ethics, Critical Thinking, K-12.}

\maketitle

\section{Introduction}
The next generation of software developers is learning programming alongside GenAI tools such as ChatGPT or Claude~\cite{maurat2025,lu2023}. This transformation extends into secondary schools, where students might encounter GenAI during their first programming experiences~\cite{zhang2023,lee2022}. 
Before we can design effective programming education for the GenAI era, we need to understand how young learners actually use such tools when programming and what their perception about \emph{ethical usage} are. This knowledge is essential for developing curricula, from schools to universities, that help students build strong foundational software engineering principles~\cite{sibia2024}.\footnote{Throughout this paper, we use AI and GenAI to refer to generative AI tools (primarily large language models) that students use for programming assistance. We use AI-assisted programming to describe activities where students utilise these tools for tasks such as code generation, debugging, or explanation.} 

While tools based on GenAI support debugging, code generation, and testing~\cite{becker2023,nguyen2024a,clarke2025}, they also challenge the development of code comprehension, frustration tolerance, and critical thinking~\cite{nguyen2024a,hashmi2024}. 
Traditional programming education tends to emphasise predominantly writing code~\cite{lachney2021,lister2004}, with evaluating solutions and code comprehension in later stages~\cite{dwyer2015,flores2012}. With the current GenAI movement, readily available tools may encourage reliance on generated solutions rather than critical assessment~\cite{clarke2025,zamfirescu-pereira2023}. Research with university students demonstrates this clear pedagogical tension: the use of AI can improve motivation and computational thinking~\cite{yilmaz2023,suriano2025}, but may also foster superficial engagement and incomplete understanding of programming concepts~\cite{dangol2025,clarke2025}.

Therefore, developing AI literacy\footnote{AI literacy refers to the knowledge and skills necessary to understand, evaluate, and use AI systems responsibly.} at an early stage of gaining programming experience is a necessity~\cite{jia2025}. Students must understand AI and its limitations, ethical and societal implications, and responsible use when creating software~\cite{ng2021,carolus2023,jia2025}, including accountability as potential future software developers~\cite{jobin2019,ng2024b,bartsch2026}. 

Despite this importance, we lack an empirical understanding of how secondary students approach AI-assisted programming. Research focuses overwhelmingly on university students~\cite{clarke2025,suriano2025}, with little examining gender differences~\cite{maurat2025}. This is concerning given documented gender disparities in traditional programming education~\cite{hsu2014,dewit2024,grassl2023e} and AI tools' potential effects~\cite{hsu2022}. Since software engineering education is situated within cultural context~\cite{lachney2021,scott2015a}, especially regarding GenAI~\cite{eguchi2024}, the German-specific GenAI discourse adds complexity: GDPR\footnote{The General Data Protection Regulation (GDPR, 2018) is the EU’s data privacy law that grants individuals strong rights and imposes strict obligations on organisations.} and EU AI Act\footnote{The EU AI Act (2024) is the world’s first comprehensive legal AI framework, classifying AI systems by risk and requiring transparency, safety, and accountability.} regulations~\cite{europeanparliamentandcouncil2024} create cultural emphasis on data protection that may shape attitudes differently than less stringent environments~\cite{custers2018,ivkovic2025}, potentially challenging international university teamwork~\cite{berrezueta-guzman2024a,grassl2023d}. Without understanding incoming students' perceptions, universities risk curricula that over-estimate technical skills or under-utilise existing ethical reasoning, potentially widening gender gaps~\cite{maurat2025}.

We address this gap through an exploratory study with 84 German secondary students (aged 16--19) interested in software engineering and computer science, recruited through extracurricular software development workshops, ensuring programming experience. We employed a mixed approach~\cite{ng2024,styve2024} to address the following research questions:

\textbf{RQ1:} How do young programming novices perceive their critical thinking in AI-assisted programming, and does it differ by gender? 

\textbf{RQ2:} How do young programming novices perceive AI ethics in AI-assisted programming, and do these differ by gender?

Our findings reveal an AI paradox: while students report sophisticated ethical reasoning about risks when using GenAI tools during programming, their responses reveal concerning gaps in practical programming discipline, particularly in their willingness to integrate incomprehensible code. We observed limited gender differences, with boys preferring AI-first problem-solving and girls emphasising collaboration with human peers.

This exploratory study lays the foundation for understanding how culture, gender, and GenAI tools shape students' current usage and readiness of AI-assisted software engineering. These findings will help to guide future curricula, e.g. code ownership.

\section{Background and Related Work}
This section reviews prior research on AI-assisted programming, AI literacy, and ethics, and presents the cultural context.

\subsection{AI-Assisted Programming in Education}
AI tools fundamentally alter how students learn programming by providing real-time support and easy access during programming~\cite{nguyen2024a}. 
So far, research shows mixed outcomes regarding attitudes and learning objectives~\cite{scholl2024}. Students gain improved computational thinking, self-efficacy, and motivation when using ChatGPT~\cite{yilmaz2023} or specific CodeFlow assistants~\cite{huang2025}, while exploring diverse approaches and learning through error analysis~\cite{becker2023,hashmi2024}. Conversely, risks include dependency without critical skills, incomplete answers, and anxiety about professional futures~\cite{yilmaz2023}. Despite few studies in programming education contexts, students' prior knowledge significantly affects their engagement with AI tools~\cite {li2025}.

With the current movement of accessible and rapid code generation through GenAI tools, programming education must shift from focusing on teaching pure coding mechanics to comprehension~\cite{rubio-manzano2025}. However, research concentrates overwhelmingly on university students. 
Secondary school approaches remain largely unexplored, particularly before entry to higher education. 
In addition, there is little research on gender-specific patterns, as prior literature on \emph{traditional} programming education suggests~\cite {hsu2014,dewit2024}. One study showed that male undergraduates, especially in their first year, used GenAI in programming significantly more than their female peers~\cite{maurat2025}. At the secondary school level, when using conversational AI in a block-based programming environment, girls improved their learning outcomes, while boys often misused the AI~\cite{hsu2022}.

\subsection{AI Literacy in (Programming) Education}
Literacy concepts have expanded beyond traditional reading and writing to encompass AI competencies necessary for navigating modern environments~\cite{carolus2023}. Early frameworks focused on basic recognition and use~\cite{kandlhofer2016}, whereas contemporary models emphasise four domains: understanding mechanisms, applying tools, critically assessing outputs, and addressing ethics~\cite{long2020b,ng2021}.

In most of the educational programs for younger learners, the material suggest to prioritise principles over technical details: how machines perceive, represent knowledge, learn, interact with humans, and affect society~\cite{touretzky2019,zhang2023}. The EU AI Act defines AI literacy as skills enabling informed deployment while understanding opportunities, risks, and potential harms~\cite{europeanparliamentandcouncil2024}. 
Similar to traditional programming education, some studies advocate starting instruction of AI and programming as early as kindergarten~\cite{su2024}.

Assessment instruments measure awareness, usage, evaluation skills, and ethics~\cite{wang2023a,carolus2023}, recognising that operating AI differs from possessing genuine literacy~\cite{audrin2022,ng2012}. 
Research with secondary school students reveals an emphasis on application and social-ethical dimensions rather than technical aspects, particularly among those with lower interest in computer science~\cite {lenke2025}.

\subsection{Ethical Dimensions and Critical Thinking}
AI Ethics encompasses responsibilities and risks in AI deployment~\cite{wang2023a}, representing core literacy for \emph{appropriate} use~\cite{wilson2018}. Global analyses identify transparency, justice, non-maleficence, responsibility, and privacy as fundamental principles, though geographic bias limits representation~\cite{jobin2019,correa2023}. 

Educational frameworks address security, social responsibility, privacy, digital relationships, and harm prevention~\cite{kim2018a}, while current approaches emphasise reliability, safety, privacy protection, accountability, transparency, awareness, and social benefit~\cite{ng2024}. Effective learning requires integrating technological understanding with moral reasoning~\cite{chai2021,jobin2019,borenstein2021,zhang2021}.

Critical thinking represents a fundamental programming education objective~\cite{flores2012,ahern2019,facione1990}, as it involves questioning assumptions, evaluating alternatives, thorough testing, and prioritising understanding over copying~\cite{dwyer2015,mcloughlin2010}. The rise and use of GenAI creates tensions as unprecedented access to solutions risks superficial engagement~\cite{zamfirescu-pereira2023} and dependency without analytical capability development. AI enables personalised learning but risks over-reliance, reduced critical thinking, and privacy concerns~\cite{vieriu2025}. Middle school students can evaluate AI beyond functional knowledge, considering personal and social issues~\cite{er2023}, which require tailored education, especially for young novices~\cite{ko2025a}.

Reviews suggest AI can enhance critical thinking~\cite{premkumar2024}, with correlations between attitudes, trust, engagement, and performance~\cite{suriano2025}. However, structured use is essential, as university students demonstrate sophisticated patterns, critically assessing AI suggestions and verifying them before implementation~\cite{clarke2025,scholl2024a}.

Overall, research overwhelmingly examines higher education contexts, leaving unexplored the development of secondary or even primary school students' critical thinking with AI assistance in general, and especially regarding programming.

\subsection{German Education and Regulation}
Education is situated within a cultural context, which provides a framework of what is taught and how it is taught~\cite{sprenger2024,scott2015a}. 
Germany's federal education system creates varied computer science landscapes across its 16 states. According to the German Informatics Society, currently 75\% of students in secondary school receive \emph{some} computing instruction, but only 6\% receive recommended volumes to be proficient in programming.\footnote{Overview of computer science courses offered in the 2024/2025 school year in the 16 German federal states: https://informatik-monitor.de/2024-25} 

The German curriculum emphasises data handling, system understanding, modelling, problem-solving, and societal evaluation. Programming typically begins at age 13 with algorithms, intensifying later with basic structures and implementation. 
In contrast, for example England's approach includes mandatory computing and programming from primary school onwards, which highlight Germany's limitations.

Germany promotes AI literacy but faces challenges: privacy concerns, gaps in teacher expertise, and infrastructure deficiencies. The regulatory environment profoundly shapes attitudes. GDPR establishes globally advanced protection standards~\cite{custers2018}, with the EU AI Act creating comprehensive regulation including strict educational application rules~\cite{europeanparliamentandcouncil2024,gstrein2024,canterogamito2024,hacker2023,ivkovic2025}. This contrasts with fragmented US approaches~\cite{onoja2025} and diverse global challenges~\cite{sharma2024}.

\textbf{Research Gap.} 
Despite growing research on AI literacy and AI-assisted programming, critical gaps remain. First, existing work overwhelmingly focuses on university students, missing the formative secondary school years when students first encounter programming alongside AI. Second, gender differences in AI-assisted programming remain largely unexplored, despite documented disparities in traditional programming education. Third, cultural context, particularly strict regulatory environments like Germany's, has received insufficient attention in understanding how students reason about AI ethics and responsibility.

Therefore, we examine German secondary students’ AI-assisted programming practices, critical thinking, and ethical literacy,  while also highlighting gender differences. Our aim is to provide  guidance on further investigation to ensure this young group is not left behind in software engineering education.

\section{Method}
The objective of this study is to explore gender-specific differences in critical thinking practices and AI literacy regarding ethics among young programming novices. 

\subsection{Study Design}
\begin{table}[t]
\centering
\small
\caption{Survey items grouped by construct and source.}
\label{tab:survey_items}
\begin{tabular}{lp{6.4cm}l}
\toprule
Var. & Question & Source \\
\midrule
\multicolumn{3}{l}{Programming and AI Tool Usage}\\
\midrule
DE01 & How did or do you learn programming outside of school contexts? & [new] \\
DE02 & Do you use GenAI tools for programming? If you do use them, for which programming tasks and how often? &  [new] \\
\midrule
\multicolumn{3}{l}{Critical Thinking Practices (RQ1)}\\
\midrule
 &\emph{The next questions are about your programming process, i.e., coding and designing (whether in Java, Python or using blocks) and your use of GenAI tools.} &\\
CT01 & I use a piece of code in my program even if I do not fully understand it. & ~\cite{styve2024} \\ %AD04\_01
CT02 & I take the time to evaluate the pros and cons of alternative solutions. & ~\cite{styve2024} \\ %AD04\_02
CT03& I check code for defects. & ~\cite{styve2024} \\ %AD04\_03 
CT04 & I should test the code even if someone else has already tested it. &  ~\cite{styve2024} \\ %AD04\_09
CT05 & I clarify my thoughts/code by explaining it to others. &  ~\cite{styve2024} \\ %AD04\_05
CT06 & If I need help, I ask the GenAI before I go to my teammate. &  ~\cite{styve2024} \\ %AD04\_13
CT07 & One source of information is enough to find a solution. & ~\cite{styve2024} \\ %AD04\_07
CT08 & It is OK to settle with the first solution (code) I can find. &  ~\cite{styve2024} \\ %AD04\_04
CT09 & If I am not sure about something, I'll let it be. &  ~\cite{styve2024} \\ %AD04\_11
CT10 & Solutions found on the Internet are trustworthy. &  ~\cite{styve2024} \\ % AD04\_10
CT11 & Solutions generated by GenAI are trustworthy. &  ~\cite{styve2024} \\ %AD04\_12
CT12 & When you use GenAI tools for programming, how do you consider ethical aspects? (Example: data privacy, what information you enter into  GenAI) & [new] \\
\midrule
\multicolumn{3}{l}{AI Ethics (RQ2)}\\
\midrule
 &\emph{The next questions are about AI ethics from your perspective as a programmer, including responsible use, fairness, accountability, and the impact of GenAI tools.} &\\
AE01 & I understand how misuse of GenAI could result in substantial risk to humans. & ~\cite{ng2024} \\ %ai01
AE02 & I think that GenAI systems need to be subjected to rigorous testing to ensure they work as expected. &  ~\cite{ng2024} \\ %AI01\_02
AE03 & I think that users are responsible for considering AI design and decision processes. &~\cite{ng2024} \\
AE04 & I think that GenAI systems should benefit everyone, regardless of physical abilities and gender. & ~\cite{ng2024} \\
AE05 & I think that users should be made aware of the purpose of the system, how it works and what limitations may be expected. & \cite{ng2024} \\
AE06 & I think that people should be accountable for using GenAI systems. & ~\cite{ng2024} \\ 
AE07 & I think that GenAI systems should meet ethical and legal standards. & ~\cite{ng2024} \\ %AI01\_07
AE08 & I think that GenAI can be used to help disadvantaged people. & ~\cite{ng2024} \\ %AI01\_08
AE09 & In your opinion, who is responsible for ensuring that ethical aspects such as data protection are taken into account in GenAI tools? & [new] \\
\bottomrule
\end{tabular}
\end{table}

We conducted a survey to capture students' experiences and perceptions regarding programming and the use of GenAI tools. 
\Cref{tab:survey_items} presents the questionnaire, which was designed to measure three thematic areas: (1) demographic and background information, (2) critical thinking practices in programming (RQ1), and (3) AI literacy regarding ethical learning (RQ2). Survey items were adapted from validated instruments from prior studies in software engineering education~\cite{styve2024} and AI education~\cite{ng2024}.   

\paragraph{Demographics}
Demographic items collected essential background information to contextualise survey responses and support analysis of gender-specific patterns. Students reported their gender, age, ethnicity, prior computer science instruction at school, programming experience, previous exposure to ethics, and self-directed learning behaviours outside of school (\emph{DE01}). 

Additionally, questions on AI tools asked which tools students use, how often they use them, and the programming tasks for which they apply these tools (\emph{DE02}). This information allows us to evaluate students’ prior experience and exposure to AI-assisted programming, as any response indicating they had not used AI assistance for programming would be excluded from the dataset.

\paragraph{Critical Thinking Practices}
Critical thinking questions were adapted from Styve et al.~\cite{styve2024}, who developed and validated 19 items for introductory university-level AI programming courses. 

We selected eight questions relevant for secondary school students, covering reflection on code, evaluation of alternatives, verification of solutions, and use of organisational tools. This was done with two external researchers from software engineering education, who have experience in conducting studies with primary and secondary school students and teaching, and who are not part of the research team. 
In addition, we created one open-ended question to explore how students consider data protection when using generative AI. 

All single-choice questions use a five-point Likert scale. Items were mapped to established critical thinking skills and sub-skills~\cite{facione1990}, such as analysis, evaluation, inference, explanation, and self-regulation. We asked students to reflect on their usual programming habits at school, at home, or in their courses to answer questions about their critical thinking practices.

\paragraph{AI Literacy: Ethical Learning}
AI literacy and ethical learning items were drawn from the AI Literacy Questionnaire (AILQ)~\cite{ng2024}, which evaluates students' literacy development across the affective, behavioural, cognitive, and ethical dimensions. This questionnaire is designed for secondary/high school students and was validated through external studies~\cite{lintner2024}. 

We use the questions from the ethical dimensions and include an additional open-ended question to capture students’ perceptions of responsibility for ensuring ethical aspects, such as data protection. The combination of structured Likert-scale items and open-ended questions provides both quantitative and qualitative insights into ethical reasoning. In the survey, we asked students to focus on their programming practices and everyday life when answering questions about AI ethics.

\subsection{Data Collection}
Data were collected during the summer of 2025 from two extracurricular settings: a local and national software development workshop targeted for secondary school students in Germany. One workshop was a one-day event, while the other lasted five days; in both contexts, students worked collaboratively in teams to develop small software projects. Those projects were, for example, small games in Java or Python, as well as app development through App Inventor.

The chosen contexts enabled us to reach a target population of students who are motivated, likely to pursue studies in computer science, and have basic programming experience. 
Given that computer science instruction is not mandatory in most German states and that school access to research is limited, these extracurricular programs provide a suitable and realistic setting for collecting relevant data. 

Instructors of the workshops did not provide guidance or recommendations regarding AI tools, ensuring that survey responses reflected students’ independent experiences. The instructors of the workshops were also not part of the research team. The survey was administered online shortly before the conclusion of the workshops using \emph{SoSci Survey}\footnote{\url{https://www.soscisurvey.de/}}. 

Before handing out the survey, we explained the study's purpose and procedure, introduced ourselves and our work at the university, and clarified the key terms used in the study. We explained the terms using simple examples and everyday language. GenAI and AI tools were described as software that can help create or check text, code, or images, such as ChatGPT or Claude, so students understood that these systems can assist them.

Participation in the study was voluntary, and consent was obtained from the students as well as the workshop instructors. All participants were informed about the purpose of the study, the planned publication of anonymised results, and their right to withdraw at any time.

\subsection{Participants}
Our study involved 84 German secondary school students aged 16 to 19 who participated in computer science outreach workshops. 
Of these, 61 attended the five-day workshop, and 23 attended the one-day workshop. The courses were only used to reach the target group who need to be familiar with programming activities and who aim to study computer science.

The gender distribution was 35.7\% girls, 64.3\% boys, with no non-binary participants. This gender ratio is relatively balanced compared to the German context, where only about 20\% of students studying computer science and a similar share of professional software developers are female (ranging from 14\% to 26\% across domains and reports).  

The average age was 17 years, with 8.3\% aged 19, 20.3\% aged 18, 35\% aged 17, and 26.3\% aged 16. All participants were living and studying in Germany at the time of data collection.  Residence was distributed across several federal states: 8 from Baden-Württemberg (9.5\%), 35 from Bavaria (41.7\%), 27 from Hesse (32.1\%), two from Saxony (2.4\%), four from North Rhine-Westphalia (4.8\%), and eight from Rhineland-Palatinate (9.5\%).  
All students self-reported not having received any formal or informal training in ethics, AI ethics, AI privacy or AI literacy.

Almost all students (89\%) reported having attended programming classes at school, mostly elective, although these courses were often introductory (e.g., Robot Karol, Scratch, or Java basics). Self-assessed programming competence in an open question was generally described as basic, with all students being confident and able to write short programs in their preferred language. By \emph{short program} we refer to a simple application, such as a basic calculator or a trivial game, where at least one loop and a conditional statement must be used. A small number of boys (14.81\%), however, indicated advanced experience, reporting more than three years of continuous programming in their free time.

\subsection{Data Analysis}
As an exploratory study, we focus on descriptive patterns rather than causal relationships. Responses are summarised using percentages and descriptive statistics.

To detect gender-dependent differences, we performed chi-square tests on categorical responses (5-point Likert scale) between boys and girls. 
Effect sizes are reported using Cramér's $V$ (for 2×5 tables), interpreted according to Cohen~\cite{cohen1960a}: negligible $<0.2$, small $0.2$--$0.3$, medium $0.3$--$0.5$, and large $>0.5$. This approach allows us to identify both statistically significant and practically meaningful differences, despite the rather small cohort.

For the open-ended question, we conducted a thematic analysis following standard procedures~\cite{braun2012}. First, two researchers familiarised themselves with the responses and carefully read all quotes. They independently generated initial codes, which were then grouped into broader themes. Since the responses were mostly short, this process was manageable and allowed for detailed coding. Interrater-reliability was assessed by comparing the two researchers’ coding, resulting in an agreement of $0.89$, which is considered excellent according to standards~\cite{landis1977}.

\subsection{Threats to Validity}
We conducted an exploratory study~\cite{runeson2012} using self-reported data from young programming learners. As with all educational studies of this kind, several validity concerns must be considered.

\textbf{Internal validity.} The data is based on motivated students from workshops, which presents self-selection bias; however, reaching out to this young target group is particularly challenging. We also rely on students’ self-reports, which may be influenced by recall bias or social desirability. Students may overestimate their ethical reasoning or critical thinking and underestimate risky programming behaviours, particularly when integrating AI-generated code without a full understanding. To mitigate this, we assured anonymity, clarified that there were no right or wrong answers, and triangulated survey responses with qualitative quotes. Students' self-reported practices may also differ from their actual behaviour. Future work should triangulate with actual programming and code review observations. 

\textbf{Construct validity.} The survey items were adapted from established instruments on critical thinking (evaluated with undergraduates) and AI attitudes (evaluated with high school students). However, our young learners may have understood some items differently than university students and the German translation from the research team. To reduce this risk, we explained key terms such as GenAI and piloted the survey with three German students.

\textbf{External validity.} Our sample included 84 high school students in Germany. Participants were self-selected volunteers from programming camps and workshops, likely more motivated and tech-savvy than the general student population. However, this is also our target group, as it reflects German students' opinions. Still, results may not generalise to other age groups, countries, or school systems. 
Gender-specific patterns should be interpreted cautiously due to small subgroup sizes and potential cultural and contextual influences. Observed differences may also reflect both socialisation and educational context. We mitigated risks of misinterpretation by reporting both descriptive and inferential statistics and triangulating quantitative results with qualitative quotes.

To support transparency and replication, we provide all survey materials and analysis scripts online and invite researchers and educators to replicate our study.\footnote{\url{https://figshare.com/s/129b3e72a072cb11737c}}

\section{Results}
This analysis examines survey responses from German secondary school students (ages 16--19), focusing on their AI-assisted programming critical thinking practices (\emph{CT}) and their AI ethics (\emph{AE}). 

\subsection{Demographics: Programming and AI Tool Usage Patterns}

\paragraph{Learning Sources Outside School (DE01).}  
Many students reported learning through YouTube or video tutorials, with 25.4\% indicating this source, predominantly boys. 20.3\% of participants used other online tutorials and websites, with a balanced gender distribution. At the same time, school instruction, such as code clubs or workshops, was frequently reported (17\%), and a small group (13.6\%) reported having frustrating programming experiences outside school, mostly girls who struggled to start or continue programming independently. Only a minority relied on books or official documentation (10.2\%, mixed gender). 

Boys demonstrated greater self-directed learning behaviours, often with long-term engagement, as reported in research~\cite{stattkus2025}, for example, stating that they had been programming for five years in their free time. In contrast, girls showed a preference for structured learning and lower overall exposure~\cite{grassl2023e}.

\paragraph{AI Tool Usage (DE02).}  
All participants indicated in the survey that they use GenAI tool for programming assistance.
Most participants (94\%), regardless of gender, primarily use ChatGPT, which aligns with findings among university students~\cite{maurat2025}. However, their common use, namely mostly for creating code (78\%) and checking for errors in their existing code (62\%, debugging), is in contrast to German university students' use of ChatGPT in programming exercises~\cite{scholl2024a} as they use it mainly for \emph{problem and conceptual understanding}. 
Other tools mentioned included Gemini (14\%), Perplexity (12\%), Claude (9\%), and specialised resources such as Cursor, Perchance, and Ecosia AI-Chat (each under 5\%). 

Usage frequency for programming tasks varied widely: 35\% reported using AI tools daily, 45\% several times per week, and 20\% occasionally (once a week). Boys were more likely to be daily users (48\% of boys vs. 21\% of girls) and tended to experiment with multiple tools, mirroring trends observed in male undergraduates~\cite{maurat2025}. Girls showed more cautious and limited engagement. Some girls described using AI tools as a source of \emph{inspiration} rather than as a means of directly following the outputs.

\subsection{RQ1: Critical Thinking Practices in AI-Assisted Programming}
We report students' perceptions of critical evaluation and quality assurance when using AI tools for programming tasks (\emph{CT01--11}). We report baseline, gender-specific perceptions and qualitative insights (\emph{CT12}).

\subsubsection{Baseline Findings}
\begin{figure}
	\centering
	\includegraphics[width=1\linewidth]{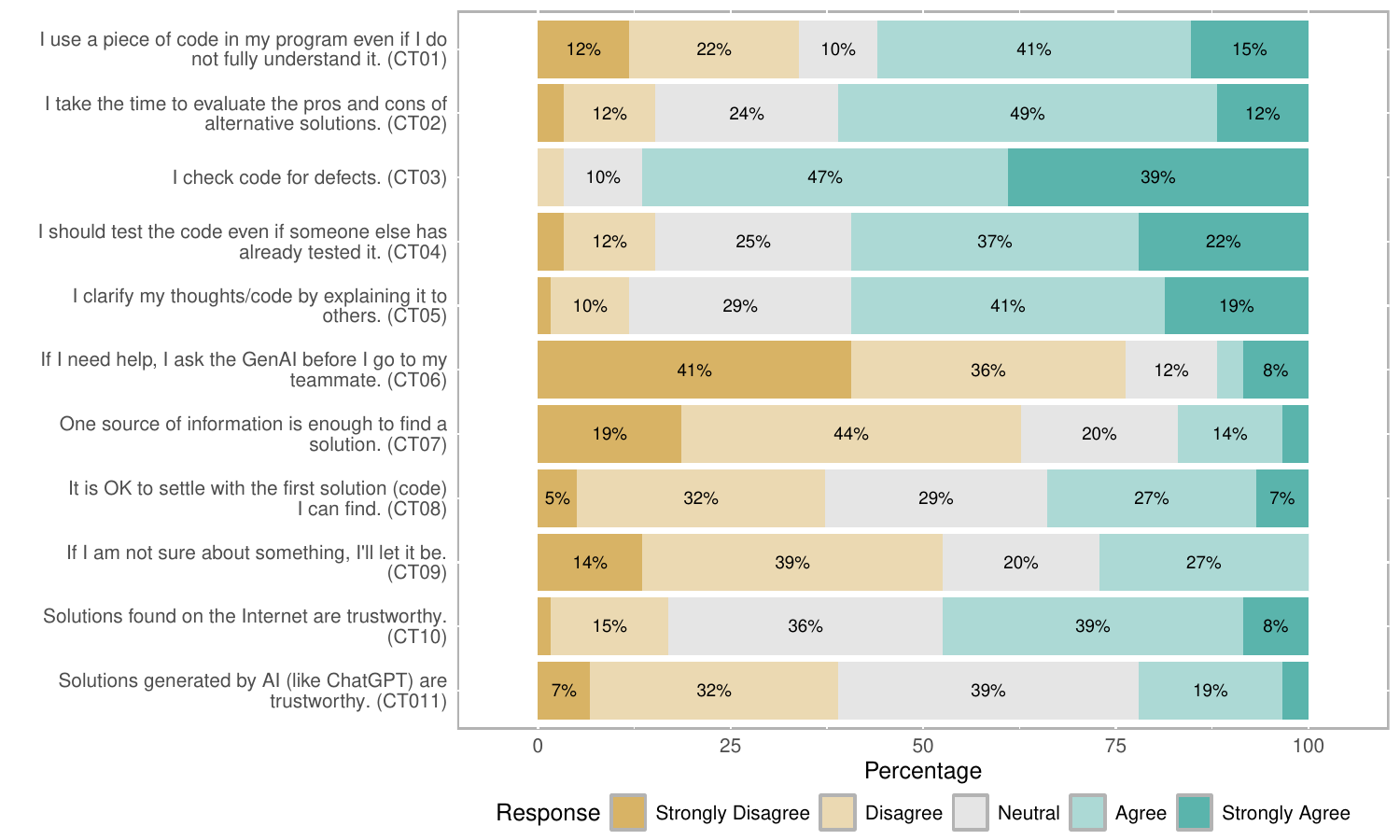}
	\caption{Students' Critical Thinking Practices.}
	\label{fig:criticiallearning}
\end{figure}

\Cref{fig:criticiallearning} shows the distribution of students' responses to the critical thinking items (\emph{CT01--11}). 

Students demonstrated strong commitment to fundamental software engineering practices, with the majority (88\%) agreeing or strongly agreeing that they should test the code even if someone else has already tested it (\emph{CT04}). Code quality checking showed similar patterns, with 86\% agreeing they check code for defects (\emph{CT03}). 
Additionally, 61\% reported taking time to consider alternative solutions (\emph{CT02}), and two-thirds disagreed that using a single source of information is sufficient (\emph{CT07}).

Regarding help-seeking behaviour, 78\% disagreed with asking GenAI before going to a person (\emph{CT06}, \Cref{fig: criticallearning}). 

However, students showed mixed feelings about poor engineering practices. Over a third of students agree and disagree with settling for the first solution of code they find (37\% disagreement, 33\% agreement, \emph{CT08}).

They also expressed mixed reactions about trust in AI-generated solutions (\emph{CT11}), 39\% of students disagreed or strongly disagreed that the solutions provided by AI tools are trustworthy. At the same time, 32\% held neutral positions and 25\% expressed agreement. 
For internet-sourced solutions, there seems to be more trust (\emph{CT10}). 

However, patterns concerning code comprehension emerged. When asked about using generated code in their program, even if they do not fully understand it (\emph{CT01}), 47\% expressed disagreement, but 32\% remained neutral, and 21\% agreed. This suggests a significant proportion may integrate incomprehensible code into their projects.

\subsubsection{Gender-specific Patterns}
\begin{figure}[t]
	\centering
	\includegraphics[width=1\linewidth]{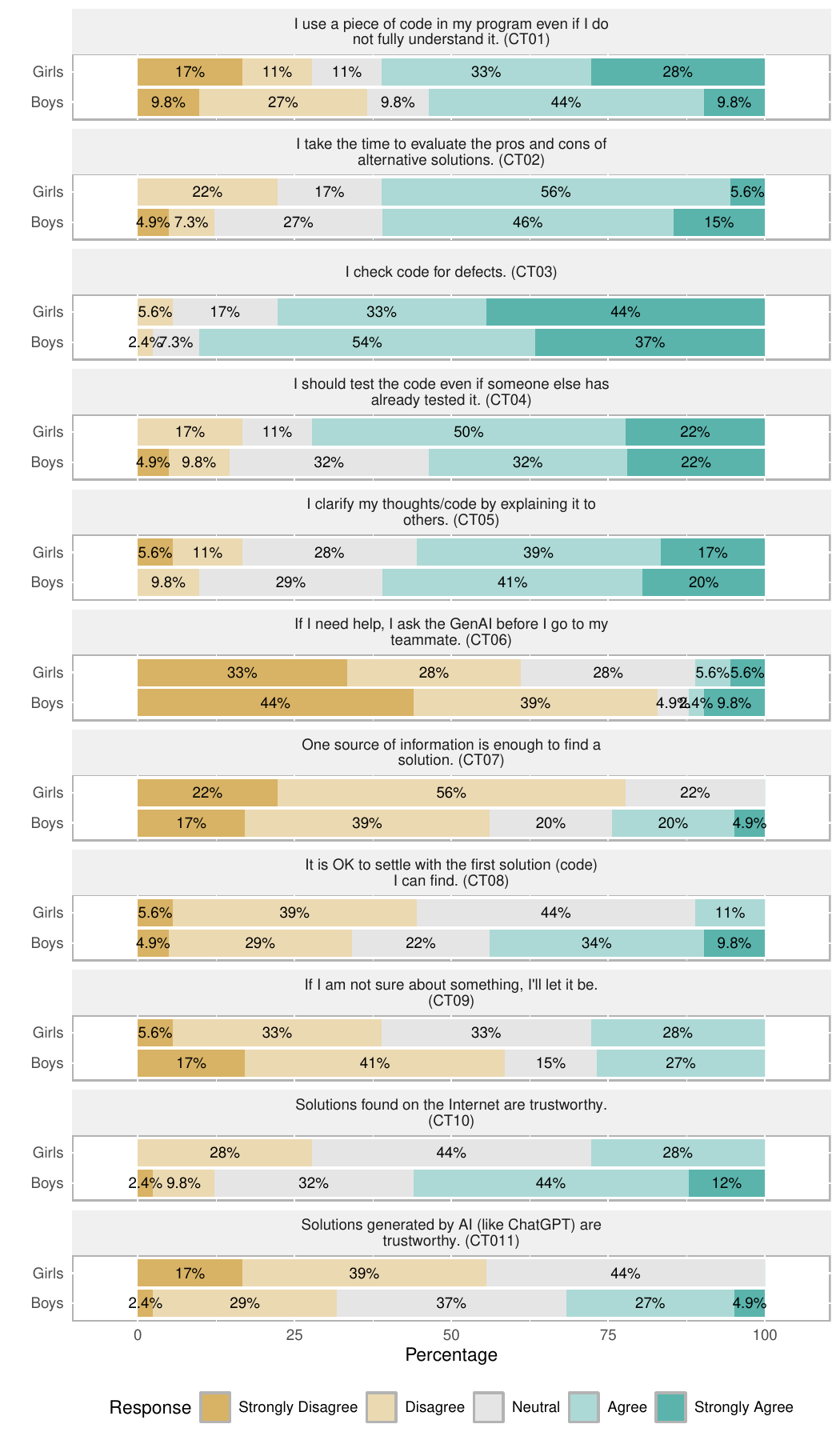}
	\caption{Gender Differences in Critical Thinking Practices.}
	\label{fig:gendercriticiallearning}
\end{figure}

\begin{table}[t]
\centering
\caption{Statistical Gender Differences for Critical Thinking Practices (RQ1).}
\label{tab:gender_differences_rq1}
\begin{tabular}{@{}p{1.8cm}rrrl@{}}
\toprule
Var. & $\chi^2$(4) & $p$-value & Cramér's $V$ & Effect Size \\
\midrule
CT01 & 6.502 & 0.165 & 0.332 & small \\
CT02 & 1.858 & 0.762 & 0.177 & negligible \\
CT03 & 1.943 & 0.746 & 0.181 & negligible \\
CT04 & 4.892 & 0.299 & 0.288 & small \\
CT05 & 4.281 & 0.369 & 0.269 & small \\
\textbf{CT06} & \textbf{10.448} & \textbf{0.034} & \textbf{0.421} & \textbf{medium} \\
CT07 & 2.983 & 0.561 & 0.225 & small \\
CT08 & 3.179 & 0.528 & 0.232 & small \\
CT09 & 3.518 & 0.475 & 0.244 & small \\
CT10 & 4.647 & 0.326 & 0.280 & small \\
CT11 & 7.291 & 0.121 & 0.351 & small \\
\bottomrule
\end{tabular}
\end{table}

\Cref{fig:gendercriticiallearning} presents the same items separated by gender (\emph{CT01--11}). 
Overall, the distribution of responses was similar across boys and girls, but several descriptive differences stand out. 

The only statistically significant difference was found in help-seeking behaviour: boys were more likely than girls to ask the GenAI before they ask a teammate (\emph{CT06}, \cref{tab:gender_differences_rq1}, $\chi^2(4) = 10.45$, $p = .034$, Cramér's $V = .42$, medium effect). This suggests that boys may show stronger AI-first preferences, whereas girls tend to consult human peers first, emphasising human collaboration. 

Beyond this, similar gender patterns emerged, while only some descriptive contrasts are noteworthy. On the item \emph{solutions generated by AI (like ChatGPT) are trustworthy}, girls expressed more scepticism: most disagreed, none agreed, while almost one third of boys agreed (31.9\%). This suggests a gendered difference in trust in AI-generated outputs, though not statistically significant. 

A similar but weaker trend appeared for trustworthy solutions on the Internet, where boys were slightly more positive and girls more cautious (\emph{CT10}). 
Similar to this, another pattern emerged for \emph{use a piece of code even if I do not fully understand it (CT01)}. Boys were more likely to agree or strongly agree, while girls were more divided between agreement and disagreement. 
By contrast, girls showed somewhat stronger endorsement of traditional quality practices such as \emph{clarifying thoughts by explaining to others (CT05} and \emph{checking code for defects (CT03)}. However, these differences were not significant.

\subsubsection{Qualitative Results.}
Thematic analysis of students’ responses (\emph{CT12}) revealed four key themes in their thinking about data privacy when using GenAI. Patterns were similar across genders.

\textbf{Privacy-Conscious Data Filtering.}
Many students, regardless of gender, reported sanitising their inputs to avoid sharing personal information. This included systematically removing names, usernames, or other identifying details and keeping the prompt as short and non-sensitive as possible. One girl explained: 
%Codes related to this theme were: \emph{Personal data avoidance, name removal, strategic sanitisation, selective disclosure}.
%Girls: 6/18 (33.3%)
%Boys: 13/41 (31.7%)
%
\begin{tcolorbox}[interviewquote]
\emph{``I always remove all names from myself or others, e.g., in an error message I always replace my username with 'User'''} ID78.
\end{tcolorbox}
%\begin{tcolorbox}[interviewquote]
%\emph{``yes, don't enter private information, always try to keep the prompt as short and non-sensitive as possible''} ID73.
%\end{tcolorbox}

\textbf{Risk-Benefit Calculation.}
Students acknowledged privacy risks but accepted them in exchange for the benefits of using GenAI. This pragmatic idea was more common among boys:
%Codes were \emph{Pragmatic acceptance, necessity justification, trust-based decisions, cost-benefit analysis} Girls: 2/18 (11.1%) Boys: 6/41 (14.6%)
%
\begin{tcolorbox}[interviewquote]
\emph{``I 'need' this tool and trust OpenAI to some extent. But I also know the risks well and have decided it benefits me more''} ID63.
\end{tcolorbox}
This trade-off is often due to the urgent need for competency: 
\begin{tcolorbox}[interviewquote]
\emph{``I do worry about that and I often think about it. However, at this moment I need the help of an “expert” more than I worry about what might be done with my data. In the future, local AI models would be more of an option for me, if they offer comparable performance.''} ID77.
\end{tcolorbox}
%Girls also weighed risks, but more cautiously. As one girl put it: 
%\begin{tcolorbox}[interviewquote]
%\emph{``Partly I give away data for school e.g. state, age etc but I always reconsider this''} ID121.
%\end{tcolorbox}

\textbf{Data Fatalism.}
A smaller group showed resignation, believing that privacy loss is unavoidable in the digital age. This position was more common among boys. For example, two boys explained:  
%Codes emerged \emph{Privacy fatalism, data inevitability, resignation, systemic acceptance}
%Girls: 1/18 (5.6%)
%Boys: 5/41 (12.2%)
\begin{tcolorbox}[interviewquote]
\emph{``I already assume that all possible data about me has been collected through other means (e.g. Google searches, social media, etc.), so I'm not afraid of it''} ID81.
\end{tcolorbox}
\begin{tcolorbox}[interviewquote]
\emph{``To me it feels like a black box: data in, data out. I don’t see how a company could really make use of my prompt.''} ID62.
\end{tcolorbox}

\textbf{Privacy Maximalism.}
Some students were really concerned, so they tried to avoid AI tools if possible, with a higher share among girls. As one girl explained: 
%Codes were \emph{Complete avoidance, privacy prioritisation, tool rejection, principle-based decisions}
%Mixed gender but slight female tendency. Girls: 2/18 (11.1%), Boys: 2/41 (4.9%)
%\begin{tcolorbox}[interviewquote]
%\emph{``I think a lot about data protection, which is why I barely use any AI tools at all''} ID42.
%\end{tcolorbox}
\begin{tcolorbox}[interviewquote]
\emph{``I do worry about that, the data could be used to train the AI further or to show me targeted advertising based on my information.''} ID110.
\end{tcolorbox}

\textbf{Minimal Concern Responses.}  
Around a quarter of students indicated little or no concern; this group was almost twice as large among boys as among girls. 

\subsubsection{Reflection on Quantitative-Qualitative Findings}
Students largely follow core software engineering practices, such as testing code and checking for defects, but show cautious trust in AI-generated solutions. Quantitative results indicate that boys are more likely to consult AI first, while girls favour human collaboration; qualitative insights align, showing that boys often accept privacy risks to use AI or exhibit fatalism about data, whereas girls are more likely to limit AI use or carefully filter inputs.

\begin{tcolorbox}[colback=white, colframe=purple!60]
\textbf{RQ1 Summary.} Most students report good practices such as defect-checking, yet many still integrate code they do not understand. Boys are more likely than girls to rely on AI tools before consulting peers. 
\end{tcolorbox}

\begin{tcolorbox}[colback=white, colframe=purple!60]
\textbf{RQ1 Interpretation.} Critical thinking is strong, but education should link principles to concrete (collaborative) programming practices to bridge the gap between awareness and action.
\end{tcolorbox}

\subsection{RQ2: AI Ethics}
We assessed students' understanding of AI ethics, accountability, and responsible deployment in software systems (\emph{AE01--08}). We report the baseline, gender specific patterns and qualitative insights (\emph{AE09}).

\subsubsection{Baseline Findings}
\begin{figure}
	\centering
	\includegraphics[width=1\linewidth]{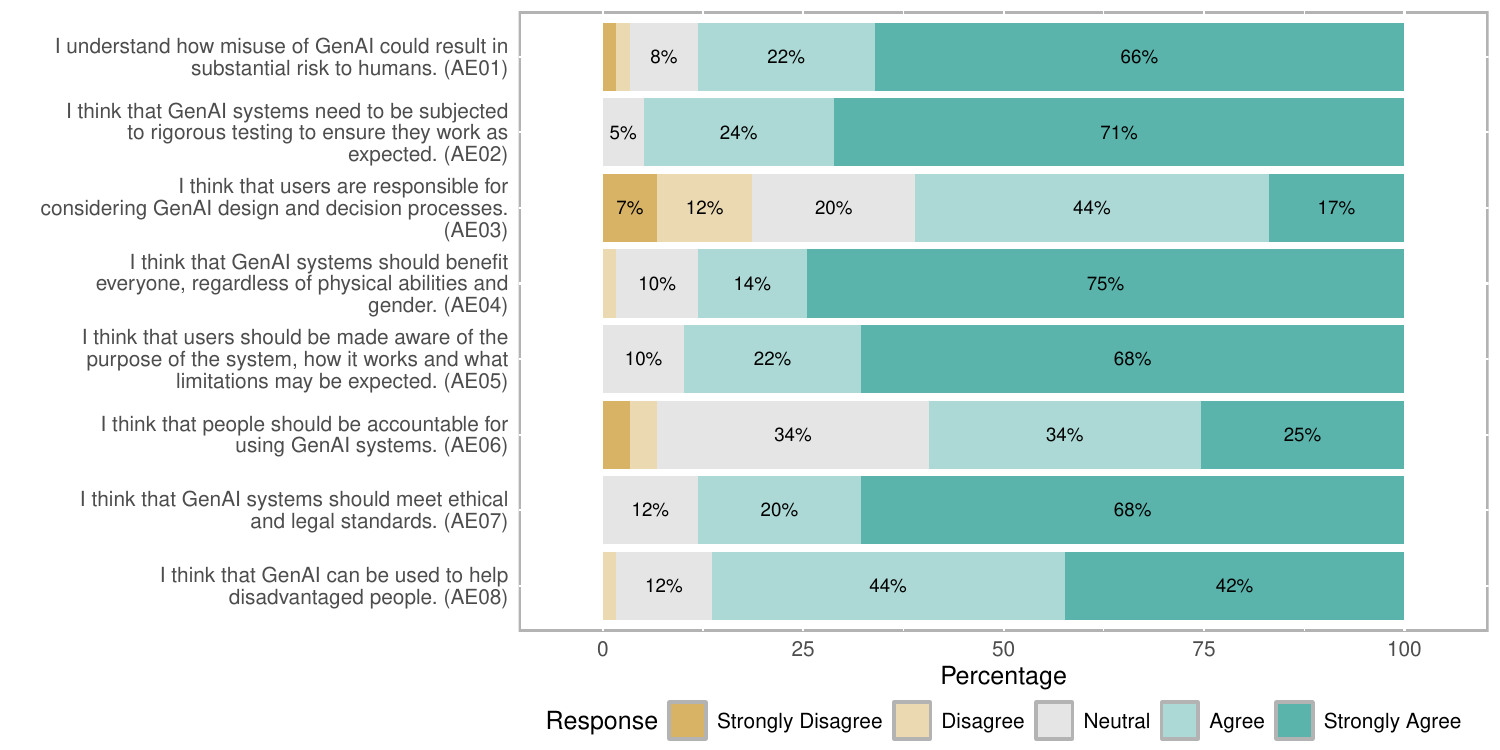}
	\caption{Students' Ethical Learning Perceptions.}
	\label{fig:ethicallearning}
\end{figure}

\Cref{fig:ethicallearning} shows the baseline results for ethical learning (\emph{AE01--08}). 
The majority of students expressed strong ethical awareness. Nearly all participants (88\%) agreed or strongly agreed that GenAI systems should meet moral and legal standards (\emph{AE07}), with none expressing disagreement. 

Transparency expectations were nearly universal, with 90\% agreeing that users should be made aware of the system's purpose, how it works, and what limitations may be expected (\emph{AE05}). Additionally, 89\% supported the principle that GenAI should benefit everyone, regardless of physical abilities and gender (\emph{AE04}).

Risk awareness was very high across the cohort, with 88\% reporting that they understand how the misuse of AI could result in substantial risk to humans (\emph{AE01}). 
Testing requirements received strong support, with 95\% agreeing that such tools need to undergo rigorous testing to ensure they work as expected (\emph{AE02}).

Regarding GenAI's social applications, 86\% agreed that it can be used to help disadvantaged people (\emph{AE08}), though this showed more variation than other ethical principles. User responsibility for AI decision processes received 61\% agreement (\emph{AE03}), indicating recognition of human agency in GenAI.

Overall, these responses suggest that students already hold a well-developed sense of AI ethics and user responsibility.

\subsubsection{Gender-specific Patterns}
\begin{figure}
	\centering
	\includegraphics[width=1\linewidth]{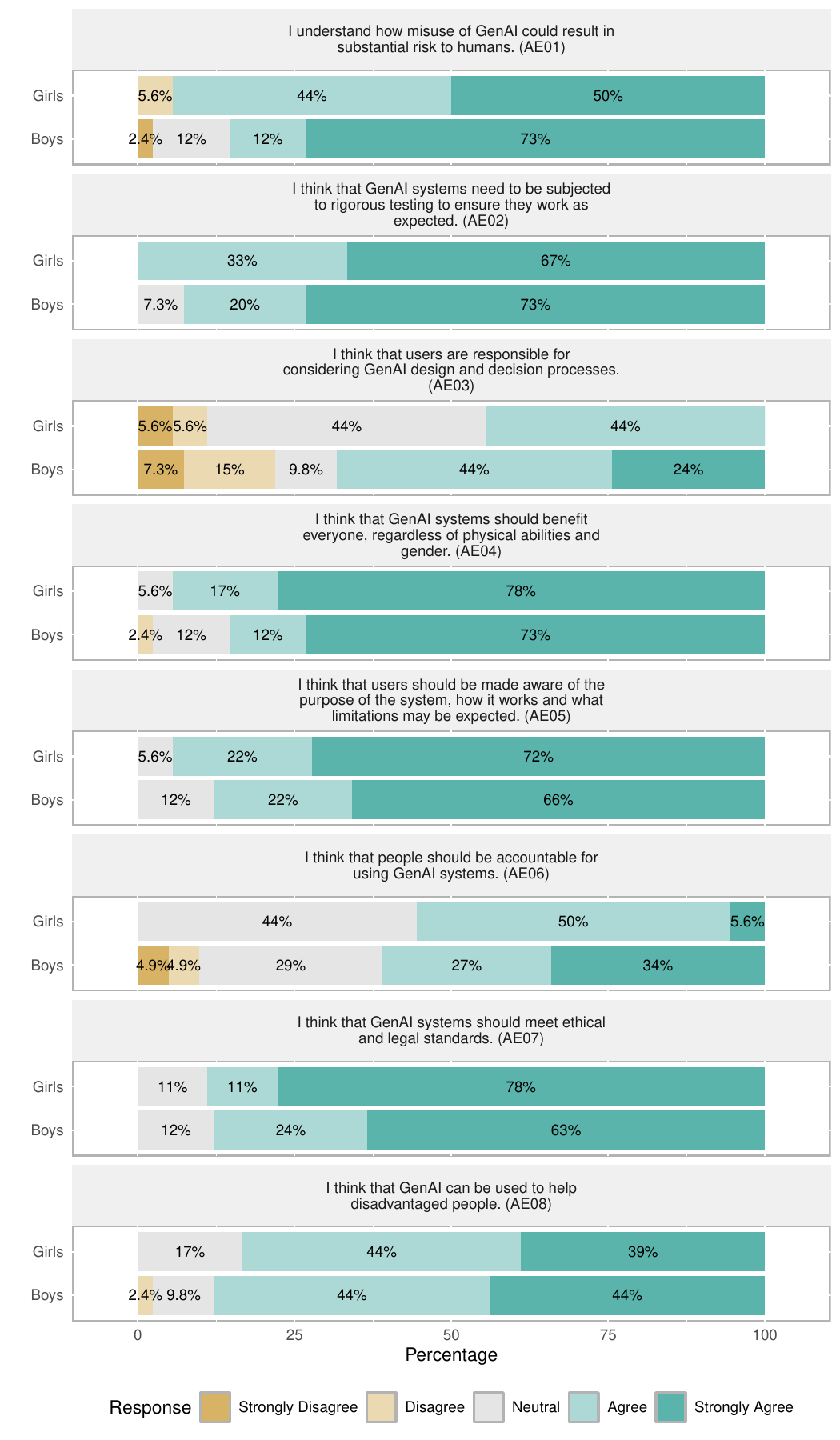}
	\caption{Gender Differences in Ethical Learning Perceptions.}
	\label{fig:genderethicallearning}
\end{figure}

\begin{table}[t]
\centering
\caption{Statistical Gender Differences for Ethical Learning (RQ2).}
\label{tab:gender_differences_rq2}
\begin{tabular}{@{}p{1.8cm}rrrl@{}}
\toprule
Var. & $\chi^2$(4) & $p$-value & Cramér's $V$ & Effect Size \\
\midrule
AE01 & 4.191 & 0.381 & 0.266 & small \\
AE02 & 2.246 & 0.691 & 0.195 & negligible \\
AE03 & 6.094 & 0.192 & 0.321 & small \\
AE04 & 3.461 & 0.484 & 0.242 & small \\
AE05 & 1.732 & 0.785 & 0.171 & negligible \\
\textbf{AE06} & \textbf{9.684} & \textbf{0.046} & \textbf{0.405} & \textbf{medium} \\
AE07 & 3.734 & 0.443 & 0.251 & small \\
AE08 & 2.913 & 0.573 & 0.222 & small \\
\bottomrule
\end{tabular}
\end{table}

\Cref{fig:genderethicallearning} presents the gender-separated results (\emph{AE01--08}). Most items showed no significant differences. 

However, chi-square analysis identified one item with a gender effect: girls were more likely to agree that \emph{people should be accountable for using AI systems (AE03)} (\cref{tab:gender_differences_rq2}, $\chi^2(4) = 9.68$, $p = .046$, Cramér's $V = .41$). 
This indicates that girls place relatively stronger emphasis on human responsibility in GenAI use. 

Overall, while consensus dominated for most items, girls appear to emphasise accountability and responsibility more strongly on the users' side. At the same time, boys tend to be more optimistic about GenAI’s potential benefits, e.g., for disadvantaged people.

\subsubsection{Qualitative Data.} 
Thematic analysis of the open question (\emph{AE09}) revealed six main themes in how students attribute responsibility for AI ethics and data privacy. Responses showed a broad consensus that regulation is essential, but differed in how responsibility should be distributed.

\textbf{Multi-Stakeholder Governance Models.}
The majority of both girls and boys stressed that politics, companies, and developers must share responsibility.
%codes emerged \emph{Shared responsibility, multi-party coordination, collaborative governance, stakeholder interdependence} Girls: 3/18 (16.7%) Boys: 9/41 (22.0%)
\begin{tcolorbox}[interviewquote]
\emph{``Politics, companies and developers. The former because they have the job of ensuring people's wellbeing, companies because they must have responsibility for their product and developers because they can ultimately prevent problematic applications from being developed''} ID64
\end{tcolorbox}

\textbf{Political Responsibility and Regulatory Necessity.}
Over half of girls and boys argued that political institutions, particularly the EU, should establish and enforce ethical frameworks. Students agreed that companies cannot be trusted to self-regulate. 
% \emph{Government responsibility, regulatory frameworks, legal requirements, policy primacy}Girls: 11/18 (61.1%) Boys: 24/41 (58.5%)

%\begin{tcolorbox}[interviewquote]
%\emph{``The legislature, ideally the EU, because companies will never do this, it's not in their interest''} ID67, boy.
%\end{tcolorbox}

\begin{tcolorbox}[interviewquote]
\emph{``Ethical problems require guidelines, and these should be defined not in terms of maximising profit, but in the interest of society. Those must come from politics so that companies pay attention to ethical aspects, because I don't think that's necessarily in their interest.''} ID43.
\end{tcolorbox}

%\begin{tcolorbox}[interviewquote]
%\emph{``In Europa sorgen die Politiker/innen der EU dafür, das Unternehmen wie Meta, OpenAI oder Apple mit underen Daten nicht machen was sie wollen''} ID91, boy.
%\end{tcolorbox}

\textbf{Corporate Implementation Responsibility.}
While politics should set the rules, almost half of students, regardless of gender, argued that companies must implement them or at least be transparent about how they use data. They were sceptical of corporate motives, stressing the need for external enforcement:  
%codes emerged \emph{Corporate implementation, profit motivation scepticism, product responsibility, data handling duties}
%In my opinion, companies are responsible because they are their tools and thus must ensure compliance with data protection' Girls: 8/18 (44.4%) Boys: 20/41 (48.8%)
\begin{tcolorbox}[interviewquote]
\emph{``In my opinion, companies are responsible because they are their tools and thus must ensure compliance with data protection. [...] They are all so money-oriented that they won't do it unless they face bigger penalties.''} ID68
\end{tcolorbox}

\textbf{Developer-Centric Responsibility.}
Several students emphasised that software developers themselves should bear responsibility. While expressed by both boys and girls, female students placed slightly more emphasis on this position.
%Overall, 9 students (15.3\%) articulated this stance, including 3 out of 18 girls (16.7\%) and 6 out of 41 boys (14.6\%).  
%Total: 9 students (15.3%) Girls: 3/18 (16.7%) Boys: 6/41 (14.6%)

\begin{tcolorbox}[interviewquote]
\emph{``Quite simply: developers, those who create the mess must also clean it up.''} ID69
\end{tcolorbox}

\textbf{Individual Agency and User Responsibility.}
Some students emphasised that users also bear responsibility, particularly in protecting their own data. This view was common among girls, while boys had strong opinions pro and con. One boy explained: 
%\emph{User agency, individual responsibility, self-protection, personal accountability} Girls: 1/18 (5.6%) Boys: 7/41 (17.1%

\begin{tcolorbox}[interviewquote]
\emph{``Realistically, the user, because all other parties never care as much about the rules as the user. Therefore, they are themselves responsible for compliance. [...] they shouldn't put private information on the internet either, AI counts as that.''} ID50,
\end{tcolorbox}

\textbf{Educational Institution Involvement.}
A minority of students highlighted the role of schools in building AI literacy and awareness of ethical issues. One girl linked politics and education: 
%Girls: 3/18 (16.7%) oys: 1/41 (2.4%) Codes emerging ware \emph{School responsibility, educational role, teacher involvement, curriculum integration}

\begin{tcolorbox}[interviewquote]
\emph{``I think this should be communicated, especially in schools. Politics, for laws and to teach children in school, for example, not necessarily to ask ChatGPT what real surname to take at a wedding''} ID43.
\end{tcolorbox}

\subsubsection{Reflection on Quantitative-Qualitative Findings}
In our cohort, students demonstrated strong awareness of AI ethics, accountability, and social responsibility. Our quantitative results show near-universal agreement on ethical standards, transparency, fairness, and risk awareness, with girls slightly more likely to emphasise human accountability. 
Our qualitative data complement this tendency: students highlighted shared responsibility across politics, companies, and developers, with girls particularly stressing developer and user accountability. Students' pragmatic scepticism of corporate self-regulation and support for political oversight align with survey findings on risk awareness. This emphasis might be due to the German context and its regulation culture. Both data sources indicate that students recognise multiple levels of responsibility and value ethical, fair, and well-regulated AI use, with subtle gendered tendencies in emphasis.

\begin{tcolorbox}[colback=white, colframe=purple!60]
\textbf{RQ2 Summary.} Students demonstrate strong ethical awareness of AI, supporting accountability, transparency, and rigorous testing. Girls place slightly more emphasis on user and developer responsibility.
\end{tcolorbox}

\begin{tcolorbox}[colback=white, colframe=purple!60]
\textbf{RQ2 Interpretation.} Students possess a strong foundation in AI ethics, suggesting that ethical awareness can be leveraged in education to reinforce responsible programming behaviours.
\end{tcolorbox}

\section{Discussion}
This study provides the first systematic analysis of how AI-native generation approaches software engineering education, examining their critical thinking practices (RQ1) and ethical awareness (RQ2). We observed promising foundations but also critical gaps, with implications for preparing the next generation of software developers both within Germany and internationally.

\subsection{The AI Paradox: Reasoning vs. Practice}
We observe a paradoxical profile by synthesising the results of RQ1 and RQ2: students demonstrate strong software engineering fundamentals (e.g., defect checking, evaluating alternatives, assessing sources) and privacy strategies, yet many integrate AI-generated code without a full understanding. The results of RQ2 show robust ethical frameworks of the young cohort: students overwhelmingly support legal standards, transparency, and rigorous testing, yet willingly accept data and privacy loss for AI assistance, suggesting data fatalism. This tension between conceptual reasoning and practical programming suggests that students can evaluate AI risks and understand ethical responsibilities, but struggle to operationalise these insights into concrete programming practices.

Gender differences nuance this paradox: boys' AI-first approaches and higher GenAI trust suggest comfort with experimental integration but potentially less critical oversight. These findings are consistent with studies indicating that men use GenAI more frequently than women in undergraduate~\cite{maurat2025}, research~\cite{tang2025}, and secondary school contexts~\cite{hsu2022}, which might be due to greater prior experience~\cite{li2025}. 
In our study, girls emphasise peer collaboration over AI assistance, suggesting accountability-driven practices that may be underutilised in AI-heavy environments. These complementary strategies could enhance team-based software engineering if deliberately leveraged in education.

\subsection{Culturally Responsive Software Engineering}
Our German students' emphasis on EU-level governance and regulatory accountability reflects their cultural context, where privacy laws are fundamental to digital citizenship. This shows that educational approaches must acknowledge cultural values~\cite{eguchi2024,sprenger2024,neumann2024}. In Germany, strict regulation leads students to view politics and law as natural arbiters of GenAI, a pattern mirrored by German university students' use of ChatGPT in introductory programming courses~\cite{scholl2024a}, as well as similar findings from Japan~\cite{eguchi2024} and Korea~\cite{ko2025a}. However, students in countries with weaker regulatory traditions may prioritise corporate decisions or individual choice over regulations.

When looking at the bigger picture, for global software engineering collaboration, curricula must balance local expectations with international teamwork preparation as student teams with diverse perspectives might be challenging~\cite{grassl2023d,earle2024a,morris2019,neumann2024}. German students' compliance intuitions and corporate scepticism can both benefit and challenge multinational teams, requiring culturally aware team formation and project management.

\subsection{Implications for Education and Practice}
Our results challenge the assumption that high technology use among young people automatically translates into strong software engineering principles~\cite{bartsch2016}. Despite strong ethical reasoning, our students report applying these principles inconsistently in practice. We identify the following main lessons: 

\textbf{Understand (in)formal learning pathways.} Students’ current GenAI practices in programming raise the question of how they acquire these approaches in the first place, since we assume such strategies are rarely taught in schools or workshops~\cite{scholl2024}. This underscores the need for grounding research with both students and educators to understand learning pathways, peer influence, and cultural factors shaping GenAI use.

\textbf{Strengthen critical thinking in AI-assisted programming.} We need to require code explanation alongside integration. For example, we could introduce code ownership checkpoints where students must explain AI-generated code to instructors or peers \emph{before} their usage or submission. In addition, we should create AI assistance logs where students document what they asked the GenAI, what they received, and what they understood or modified. This will support professional accountability principles~\cite{bartsch2026}. 

\textbf{Connect ethical awareness to concrete practice.} One reason why such an AI paradox appears in this young cohort might be the great abstraction level of both ethical and software engineering principles. Thus, we should link ethical frameworks to everyday and concrete programming tasks by e.g. responsible prompt design, privacy-conscious workflows, and collaborative decision-making in actual programming exercises. 

\textbf{Leverage complementary learning strategies.} According to our cohort, boys' experimental GenAI use and girls' collaborative, privacy-focused approach suggest complementary strengths for team-based projects. In order to not neglect the (human) collaborative aspects of software engineering, we need to foster communication practises and social skills. However, our small sample requires further research to confirm these patterns and support all genders in their first programming experiences.

Our findings emphasise the importance of integrating AI literacy with education in regulatory environments, particularly in countries like Germany. Additionally, we suggest that cultural context plays a significant role in shaping both ethical decision-making and programming practices, which are important factors for international software engineering education~\cite{sprenger2024}.

\section{Conclusions and Future Work}
This exploratory study provided the first systematic examination of  German secondary school students' critical thinking practices and ethical perceptions in AI-assisted programming, establishing baseline data for this under-studied population. We identified an AI paradox where strong ethical reasoning coexists with risky programming practices, suggesting that awareness alone is insufficient without explicit integration into coding workflows. We revealed gender-related patterns in GenAI tool usage, collaboration preferences, and privacy concerns that might have implications for inclusive software engineering education. 

Therefore, we recommend further studies across countries with different regulatory environments to clarify how cultural context shapes ethical reasoning and programming behaviours. We also need to examine how students develop GenAI practices outside of formal teaching and how educators themselves experience and frame these tools. As a follow-up on this study, we aim to explore targeted interventions to strengthen AI-critical thinking during programming, including structured guidance on code comprehension and connecting ethical principles to concrete programming practice. Finally, putting this research into a gender- and cultural-responsive context may help foster equitable development of AI literacy and programming skills.

\section*{Acknowledgements}
We sincerely thank all students who participated in this study for their time, trust, and honest responses. We also gratefully acknowledge Emily Vorderwülbeke for sharing the initial idea of this study and providing valuable guidance in refining the research design.

\bibliographystyle{ACM-Reference-Format}
\bibliography{references}

\end{document}